\documentclass[twocolumn,showpacs,preprintnumbers,amsmath,amssymb,aps,prc,superscriptaddress]{revtex4-1}
\usepackage{graphicx}% Include figure files
\usepackage{dcolumn}% Align table columns on decimal point
\usepackage{bm}% bold math
\usepackage{hyperref}% add hypertext capabilities
\usepackage{breakurl}
\usepackage{latexsym}
\usepackage{color}

\begin{document}

%\preprint{}

\title{High-precision mass measurements for the isobaric multiplet mass equation at $A=52$}



\author{D.A.~Nesterenko}
\email[]{dmitrii.nesterenko@jyu.fi}
\affiliation{University of Jyvaskyla, P.O. Box 35, FI-40014 University of Jyvaskyla, Finland}
\author{A.~Kankainen}
\affiliation{University of Jyvaskyla, P.O. Box 35, FI-40014 University of Jyvaskyla, Finland}
\author{L.~Canete}
\affiliation{University of Jyvaskyla, P.O. Box 35, FI-40014 University of Jyvaskyla, Finland}
\author{M.~Block}
\affiliation{GSI Helmholzzentrum f\"ur Schwerionenforschung GmbH, D-64291 Darmstadt, Germany}
\affiliation{Helmholtz Institute Mainz, D-55099 Mainz, Germany}
\affiliation{Johannes Gutenberg-Universit\"at Mainz, D-55099 Mainz, Germany}
\author{D.~Cox}
\affiliation{University of Jyvaskyla, P.O. Box 35, FI-40014 University of Jyvaskyla, Finland}
\author{T.~Eronen}
\affiliation{University of Jyvaskyla, P.O. Box 35, FI-40014 University of Jyvaskyla, Finland}
\author{C.~Fahlander}
\affiliation{Department of Physics, Lund University, S-22100 Lund, Sweden}
\author{U.~Forsberg}
\affiliation{Department of Physics, Lund University, S-22100 Lund, Sweden}
\author{J.~Gerl}
\affiliation{GSI Helmholzzentrum f\"ur Schwerionenforschung GmbH, D-64291 Darmstadt, Germany}
\author{P.~Golubev}
\affiliation{Department of Physics, Lund University, S-22100 Lund, Sweden}
\author{J.~Hakala}
\affiliation{University of Jyvaskyla, P.O. Box 35, FI-40014 University of Jyvaskyla, Finland}
\author{A.~Jokinen}
\affiliation{University of Jyvaskyla, P.O. Box 35, FI-40014 University of Jyvaskyla, Finland}
\author{V.S.~Kolhinen}
\affiliation{University of Jyvaskyla, P.O. Box 35, FI-40014 University of Jyvaskyla, Finland}
\author{J.~Koponen}
\affiliation{University of Jyvaskyla, P.O. Box 35, FI-40014 University of Jyvaskyla, Finland}
\author{N.~Lalovi\'c}
\affiliation{Department of Physics, Lund University, S-22100 Lund, Sweden}
\author{Ch.~Lorenz}
\affiliation{Department of Physics, Lund University, S-22100 Lund, Sweden}
\author{I.D.~Moore}
\affiliation{University of Jyvaskyla, P.O. Box 35, FI-40014 University of Jyvaskyla, Finland}
\author{P.~Papadakis}
\affiliation{University of Jyvaskyla, P.O. Box 35, FI-40014 University of Jyvaskyla, Finland}
\author{J.~Reinikainen}
\affiliation{University of Jyvaskyla, P.O. Box 35, FI-40014 University of Jyvaskyla, Finland}
\author{S.~Rinta-Antila}
\affiliation{University of Jyvaskyla, P.O. Box 35, FI-40014 University of Jyvaskyla, Finland}
\author{D.~Rudolph}
\affiliation{Department of Physics, Lund University, S-22100 Lund, Sweden}
\author{L.G.~Sarmiento}
\affiliation{Department of Physics, Lund University, S-22100 Lund, Sweden}
\author{A.~Voss}
\affiliation{University of Jyvaskyla, P.O. Box 35, FI-40014 University of Jyvaskyla, Finland}
\author{J.~\"Ayst\"o}
\affiliation{University of Jyvaskyla, P.O. Box 35, FI-40014 University of Jyvaskyla, Finland}
\affiliation{Helsinki Institute of Physics, P.O. Box 64, FI-00014 University of Helsinki, Finland}

%\thanks{}


\date{\today}

\begin{abstract}
Masses of $^{52}$Co, $^{52}$Co$^m$, $^{52}$Fe, $^{52}$Fe$^m$, and $^{52}$Mn have been measured with the JYFLTRAP double Penning trap mass spectrometer. Of these, $^{52}$Co and $^{52}$Co$^m$ have been experimentally determined for the first time and found to be more bound than predicted by extrapolations. The isobaric multiplet mass equation for the $T=2$ quintet at $A=52$ has been studied employing the new mass values. No significant breakdown (beyond the $3\sigma$ level) of the quadratic form of the IMME was observed ($\chi^2/n=2.4$). The cubic coefficient was 6.0(32) keV ($\chi^2/n=1.1$). The excitation energies for the isomer and the $T=2$ isobaric analogue state in $^{52}$Co have been determined to be 374(13) keV and 2922(13) keV, respectively. The $Q$ value for the proton decay from the $19/2^-$ isomer in $^{53}$Co has been determined with an unprecedented precision, $Q_{p} = 1558.8(17)$ keV. The proton separation energies of $^{52}$Co and $^{53}$Ni relevant for the astrophysical rapid proton capture process have been experimentally determined for the first time.
\end{abstract}

%\keywords{}

\maketitle


\section{Introduction\label{sec:intro}}

Assuming a charge-independent nuclear force, the isobaric analogue states (IAS) in an isobaric multiplet are degenerate. Their mass differences are due to Coulomb interaction and the neutron-proton mass difference. According to the Isobaric Multiplet Mass Equation (IMME)~\cite{Weinberg1959}, the masses of IASs in a mass multiplet with an atomic mass number $A$ and isospin $T$ should lie on a parabola:
\begin{equation} 
\label{eq:imme}
M(A,T,T_Z) = a(A,T) + b(A,T)T_Z + c(A, T)T_Z^2,
\end{equation}
where the coefficients $a$, $b$ and $c$ are interpreted as being the scalar, vector and tensor Coulomb energies. High-precision Penning-trap mass measurements have offered new possibilities to investigate the validity of the IMME, and have revealed a breakdown in the quadratic form of the IMME in a few cases, such as for $A=8$ \cite{Charity2011}, $A=9$ \cite{Brodeur2012}, $A=21$ \cite{Gallant2014}, $A=31$ \cite{Kankainen2016}, $A=32$ \cite{Kankainen2010b, Kwiatkowski2009}, and $A=35$ \cite{Yazidjian2007}. In general, however, the IMME seems to describe well the masses of isospin multiplets, and it has therefore been widely used to predict the masses of the most exotic members of the multiplets.

Sometimes the quadratic form of the IMME (Eq.~\ref{eq:imme}) is not sufficient to describe the masses in an isobaric multiplet but a cubic ($dT_Z^3$) or even a quartic coefficient ($eT_Z^4$) is required. The $T=3/2$ quartets have shown an interesting, increasing trend in the cubic IMME coefficients when entering into the $f_{7/2}$ shell \cite{MacCormick2014,MacCormick2014err}. On the other hand, the quadratic IMME at $A=53$ has been recently revalidated with a reduced $\chi^2$ of 1.34, and the cubic coefficient has been found to be rather small, $d=5.4(46)$ keV \cite{Su2016} compared to $d=39(11)$ keV obtained in Ref. \cite{Zhang2012}. 

The $T=2$ quintets have not been experimentally explored in the heavier mass region but could provide further insight into the possible trend in the cubic coefficients. In this paper, we have experimentally determined the masses for $^{52}$Co, $^{52}$Fe, and $^{52}$Mn, which are members of the $T=2$ isobaric quintet at $A=52$ together with $^{52}$Cr and $^{52}$Ni. Previous IMME evaluations for the quintet have suggested that a large non-zero cubic coefficient, $d=28.8(45)$~keV, might be required for the IMME \cite{MacCormick2014, MacCormick2014err}. However, the test was not very stringent due to the lack of experimental mass values for $^{52}$Co and $^{52}$Ni. Thus, the mass of $^{52}$Co, determined here experimentally for the first time, is pivotal for testing the IMME and investigating whether there is a trend towards larger cubic coefficients for nuclei in the $f_{7/2}$ shell forming $T=2$ quintets.

In addition to the ground states of $^{52}$Co, $^{52}$Fe, $^{52}$Mn, we have studied isomeric states in $^{52}$Co and $^{52}$Fe as summarized in Table~\ref{tab:properties}. The isomeric state of $^{52}$Co is of special interest because it can be used to determine the mass of the $T=2$ IAS in $^{52}$Co. The current knowledge of the $T=2$ IAS in $^{52}$Co is based on $\beta$-decay studies of $^{52}$Ni \cite{Dossat2007, Orrigo2016}. Two prominent $\beta$-delayed proton groups with center-of-mass energies of 1057(11) keV and 1349(10) keV \cite{Dossat2007} have been observed from the IAS. Similar proton energies at 1048(10) keV and 1352(10) keV have been determined in a more recent work \cite{Orrigo2016}. The proton peaks have been attributed to the decay of the IAS to the ground state and first excited states in $^{51}$Fe known from in-beam $\gamma$-ray spectroscopy \cite{Ekman2000,Bentley2000}. The excitation energy of the IAS can thus be determined as a sum of the observed proton energy and the proton separation energy of $^{52}$Co. On the other hand, the excitation energy of the IAS can be derived from the observed $\gamma$-$\gamma$cascade ($E_{\gamma 1}=2418.3(3)$ keV, $E_{\gamma 2}=142.3(1)$ keV \cite{Dossat2007}) from the IAS to the presumed $\beta$-decaying $2^+$ isomer in $^{52}$Co. However, a discrepancy was found between the IAS energies of $^{52}$Co derived from the proton and $\gamma$-decay data when tabulated mass values were applied for $^{52}$Co and $^{52}$Co$^m$ in Ref.~\cite{Dossat2007}. The $\gamma$-$\gamma$-cascade to the isomeric state in $^{52}$Co resulted in an IAS about 600 keV higher than the proton data leading to the ground state of $^{51}$Fe. Therefore, it was proposed \cite{Dossat2007} that the ground state mass excess of $^{52}$Co might be too high in the Atomic Mass Evaluation~\cite{AME2003}. With our direct mass measurements of $^{52}$Co and $^{52}$Co$^m$, we can now determine the excitation energy of the isomeric state, and therefore infer the excitation energy of the $T=2$ IAS.

\begin{table}%[H] add [H] placement to break table across pages
\caption{Properties of the nuclides studied in this work. $T_{1/2}$ is the half-life, $I^\pi$ the spin-parity and $E_x$ the excitation energy of the isomeric state. The values estimated from isospin symmetry or from systematic trends from neighboring nuclides with the same $Z$ and $N$ parities are marked by “\#”. The values are based on Ref.~\cite{NDS52-2015} unless stated otherwise. \label{tab:properties}}
\begin{ruledtabular}
\begin{tabular}{llllllll}
Nuclide & $T_{1/2}$ & $I^\pi$ & $E_x$ (keV)\\
\hline
$^{52}$Co 		& 104(7) ms 	& $6^+\#$ &  &\\
$^{52}$Co$^m$ & 104(11) ms \cite{Hagberg1997}\footnotemark[1] & $2^+$\# & 380(100)\# \cite{NUBASE12} & \\
$^{52}$Fe 		& 8.275(8) h 	& $0^+$ &  & \\
$^{52}$Fe$^m$ & 45.9(6) d 	& $12^+$ & 6958.0(4) &  \\
$^{52}$Mn 		& 5.591(3) d 	& $6^+$ &  & \\
\end{tabular}
\footnotetext[1]{Authors in Ref.~\cite{Hagberg1997} do not have specific evidence for $^{52}$Co$^m$.}
\end{ruledtabular}
\end{table}

The masses of $^{52}$Co and $^{52}$Fe discussed in this paper were measured in conjunction with a post-trap spectroscopy experiment dedicated to the study of proton radioactivity from the $19/2^-$ isomer in $^{53}$Co. It is the isomer from which the first observations of proton radioactivity were made about 45 years ago \cite{Jackson1970, Cerny1970, Cerny1972}. In this respect, the mass of $^{52}$Fe is important as when combined with the former, precise mass measurements of $^{53}$Co and $^{53}$Co$^m$ \cite{Kankainen2010a}, as it provides a precise, external calibration point for proton-decay spectroscopy. 

The nuclei studied in this work are also relevant for studies of the astrophysical rapid proton capture ($rp$) process occurring, for example, in type I $X$-ray bursts \cite{Wallace1981, Schatz1998}. The proton capture rates as well as their inverse photodisintegration reactions depend sensitively on the reaction $Q$ values \cite{Parikh2009}. In particular, the ratios of $^{51}$Fe$(p,\gamma)^{52}$Co-$^{52}$Co$(\gamma,p)^{51}$Fe and $^{52}$Co$(p,\gamma)^{53}$Ni-$^{53}$Ni$(\gamma,p)^{52}$Co reactions affecting the route towards heavier elements have been studied with experimental $Q$ values for the first time.

\section{Experimental Method\label{exp}}
The isotopes of interest were studied in two separate experiments at the Ion-Guide Isotope Separator On-Line (IGISOL) facility \cite{Moore2013}: $^{52}$Co and $^{52}$Fe in April and $^{52}$Mn in August 2015. A 50-MeV proton beam from the K-130 cyclotron impinging into an enriched 1.8-mg/cm$^2$-thick $^{54}$Fe target was used to produce $^{52}$Co and $^{52}$Fe via fusion-evaporation reactions, whereas for $^{52}$Mn a 40-MeV proton beam was applied. The reaction products were stopped in helium gas, extracted and guided towards the mass separator using a sextupole ion guide (SPIG) \cite{Karvonen2008} before acceleration to 30 kV. A good fraction of the ions are singly-charged, and the mass number $A=52$ could be selected using a $55^\circ$ dipole magnet. A gas-filled radio-frequency quadrupole cooler and buncher \cite{Nieminen2001} cooled the ions and converted the continuous beam into narrow ion bunches which were injected into the JYFLTRAP double-Penning-trap mass spectrometer \cite{Eronen2012}. In the first trap, ions were cooled, centered and purified via a mass-selective buffer gas cooling technique \cite{Savard1991}. The masses of ions with charge-to-mass ratio $q/m$ were measured in the second measurement trap by determining their cyclotron frequency $\nu_c=qB/(2\pi m)$ in a magnetic field strength $B$ via a time-of-flight ion cyclotron resonance (TOF-ICR) technique \cite{Graff1980,Konig1995}. 

The measurements of the ions of interest were sandwiched by similar measurements of the reference ion $^{52}$Cr$^+$, which were linearly interpolated to the time of the actual measurement of the ions of interest to determine the magnetic field strength. The atomic masses were derived from the cyclotron frequency ratio $r=\nu_{c,ref}/\nu_{c}$ between the reference ion $^{52}$Cr$^+$ and the ion of interest via $m = r\cdot(m(^{52}$Cr$)-m_e)+ m_e$. 

Ion-ion interactions were studied by performing count-rate class analysis \cite{Kellerbauer2003} for the determined frequencies, except for $^{52}$Co and $^{52}$Co$^m$, for which it was not necessary since most of the bunches contained only one ion. No significant differences were observed when the count-rate class analysed frequency ratios were compared with the results obtained by restricting the number of ions to one to five ions/bunch. Thus, limiting the number of ions to one to five ions/bunch is sufficient to avoid possible frequency-ratio shifts due to ion-ion interactions for the present uncertainty level.

\begin{figure}[tbh]
\centering
\includegraphics[width=0.45\textwidth,clip]{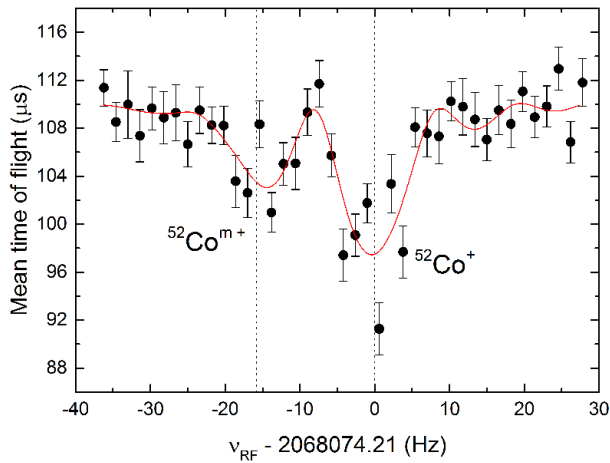}
\caption{(Color online) Time-of-flight ion cyclotron resonance spectrum for $^{52}$Co$^+$ and $^{52}$Co$^{m+}$ with a 100 ms RF excitation time. The solid red line is a fit of the theoretical curve to the data points.\label{fig:co52}}
\end{figure}

The uncertainties due to temporal fluctuations in the magnetic field, $\delta B/B=8.18(19)\times 10^{-12}$/min$ \times \Delta t$[min] \cite{Canete2016} were negligible compared with the statistical uncertainties of the measured frequency ratios. For mass doublets with the same $A/q$, the mass-dependent and systematic uncertainties resulting from field imperfections cancel in the frequency ratio \cite{Roux2013}. The internal and external uncertainties of the measured frequency ratios \cite{Birge1932} were compared and the ratio was found to be close to unity. The larger of the two values was used for the weighted mean of the frequency ratios.

For $^{52}$Co, the trap cycle was kept as short as possible due to the short half-lives of $^{52}$Co$^+$ and $^{52}$Co$^{m+}$ (see Table~\ref{tab:properties}). A single, 100-ms-long quadrupolar radiofrequency (RF) excitation applied in the second trap was sufficient to resolve the ground-state $^{52}$Co$^+$ and isomeric-state $^{52}$Co$^{m+}$ ions as shown in Fig.~\ref{fig:co52}. Although each $^{52}$Co measurement took around three hours, the uncertainty due to temporal fluctuations in the magnetic field was still much less than the statistical uncertainty of the frequency ratio ($\approx$1-2$\times10^{-7}$).

\begin{figure*}[tbh]
\centering
\includegraphics[width=0.95\textwidth,clip]{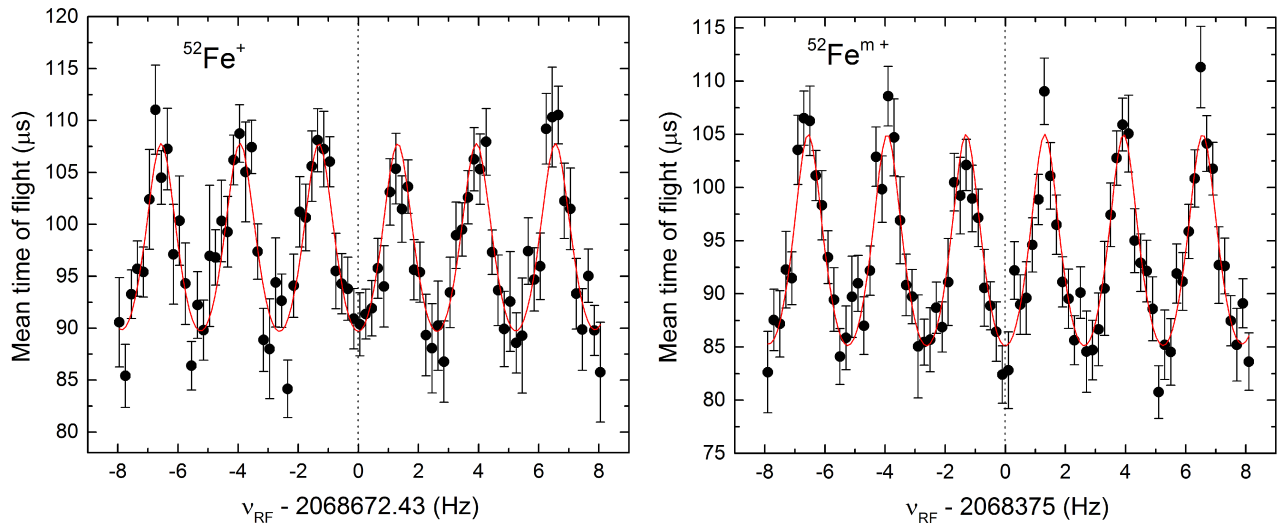}
\caption{(Color online) Time-of-flight ion cyclotron resonances of $^{52}$Fe$^+$ (left) and $^{52}$Fe$^{m+}$ (right) with 25 ms (On) - 350 ms (Off) - 25 ms (On) Ramsey excitation pattern. The solid red line is a fit of the theoretical curve to the data points.\label{fig:fe52}}
\end{figure*}

For $^{52}$Fe$^+$, $^{52}$Fe$^{m+}$, and $^{52}$Mn$^+$, the following Ramsey excitation patterns \cite{Kretzschmar2007, George2007} were applied: 25 ms (On) - 350 ms (Off) - 25 ms (On) for $^{52}$Fe$^+$ and $^{52}$Fe$^{m+}$, and 25 ms (On) - 750 ms (Off) - 25 ms (On) for $^{52}$Mn$^+$. The data for these nuclides were collected interleavedly \cite{Eronen2009}: after one frequency scan for the reference ion, a few frequency scans were collected for the ions of interest. This pattern was repeated as long as required for sufficient statistics (typically for a few hours). Such interleaved scanning reduces the uncertainties due to time-dependent fluctuations in the magnetic field considerably. Because of the high excitation energy of the isomeric state $^{52}$Fe$^m$ ($E_x=6958.0(4)$ keV \cite{NDS52-2015}), it was possible to separate the ground and isomeric states already in the first trap and measure them separately in the second trap. Examples of TOF-ICR resonances for $^{52}$Fe$^+$ and $^{52}$Fe$^{m+}$ are given in Fig.~\ref{fig:fe52}.

For the cyclotron frequency measurements of $^{52}$Mn, a Ramsey cleaning technique \cite{Eronen2008} was additionally applied to resolve the $6^+$ ground state and the $2^+$ isomeric state with excitation energy 377.749(5) keV \cite{NDS52-2015}. A dipolar excitation pulse with a Ramsey pattern 5 ms (On) - 25 ms (Off) - 5 ms (On) in the second trap excited the motion at the modified cyclotron frequency of unwanted, isomeric-state ions, but the ions of the $^{52}$Mn ground state were unaffected. Following this excitation step, only the ground-state $^{52}$Mn$^+$ ions could pass through the 1.5-mm diaphragm back to the first trap for recooling and recentering before the actual mass measurement in the second trap.

\section{Results and Discussion\label{sec:results}}
The results of the mass measurements are summarized in Table~\ref{tab:results}. Detailed discussion related to the masses of $^{52}$Co, $^{52}$Fe and $^{52}$Mn can be found from sections~\ref{sec:52co}, \ref{sec:52fe}, and \ref{sec:52mn}, respectively. In addition, the results for the excitation energies of the isomer $^{52}$Co$^m$ and the $T=2$ IAS in $^{52}$Co are discussed in section \ref{sec:52coIAS}. The impact on the proton separation energy of $^{53}$Co is explored in section~\ref{sec:Sp53Co}. In section~\ref{sec:IMME}, the IMME for the $T=2$ quintet at $A=52$ is studied in detail. Implications for the rapid proton capture process are briefly discussed in section~\ref{sec:rp}.

\begin{table*}%[H] add [H] placement to break table across pages
\caption{The weighted average cyclotron frequency ratios, $r$, and mass-excess values, $ME_{JYFL}$, determined in this work in comparison with the mass-excess values from AME12 \cite{AME12}. The atomic mass excess value -55416.1(6) keV for $^{52}$Cr \cite{AME12} was taken to calculate the mass excesses of the studied nuclides from the frequency ratios. The mass-excess values from AME12 \cite{AME12} are given in the fourth column. The differences between the JYFLTRAP and the AME12 mass values are given in the fifth column.\label{tab:results}}
\begin{ruledtabular}
\begin{tabular}{lllll}
Nuclide & $r$ & $ME_{JYFL}$ (keV) & $ME_{AME12}$ (keV) & JYFL-AME12 (keV)\\
\hline
$^{52}$Co 		& 1.00043584(14) 		& -34331.6(66) 	& -33990(200)\# & -342(200) \\
$^{52}$Co$^m$ & 1.00044356(23) 		& -33958(11) 		& -33610(220)\# & -348(220) \\
$^{52}$Fe 		& 1.0001464894(28)  & -48330.67(60) & -48332(7) & 1(7) \\
$^{52}$Fe$^m$ & 1.0002903590(56)  & -41370.01(65) & -41374(7) & 4(7) \\
$^{52}$Mn 		& 1.0000973119(13)  & -50709.97(59) & -50706.9(19) & -3(2) \\
\end{tabular}
\end{ruledtabular}
\end{table*}

\subsection{The masses of $^{52}$Co and $^{52}$Co$^m$\label{sec:52co}}
\begin{figure*}[tbh]
\centering
\includegraphics[width=0.95\textwidth,clip]{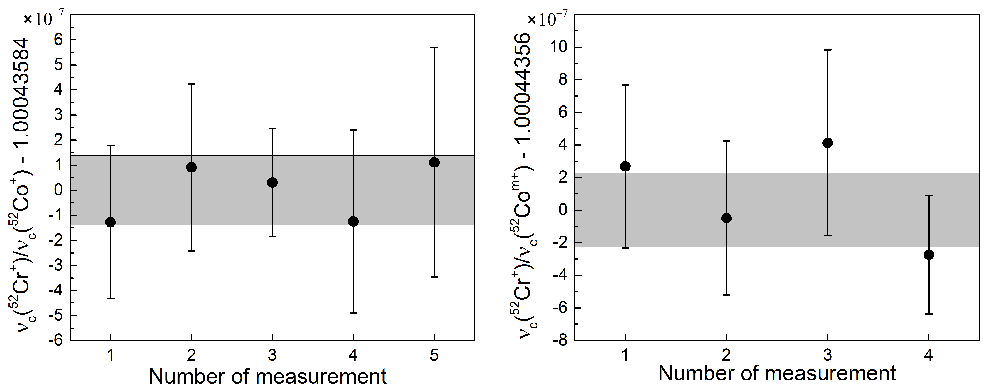}
\caption{Cyclotron-frequency ratios measured for $^{52}$Co (left) and $^{52}$Co$^m$ (right) in this work. The gray-shaded bands represent the total uncertainty of the averaged frequency ratio. \label{fig:r52co}}
\end{figure*}

The mass of $^{52}$Co has never been measured before: only an extrapolated value was given for it in the Atomic Mass Evaluation 2012 (AME12) \cite{AME12}. In this work, five cyclotron frequency ratios for the ground state and four ratios for the isomeric state were determined (see Fig.~\ref{fig:r52co}). The weighted means of the frequency ratios are $1.00043584(14)$ and $1.00044356(23)$ for $^{52}$Co$^+$ and $^{52}$Co$^{m+}$, respectively. The first experimental mass-excess values for $^{52}$Co and $^{52}$Co$^m$, $-34331.6(66)$ keV and $-33958(11)$ keV, are 342(200) keV and 348(220) keV lower than the extrapolated values in the AME12, respectively. Thus, $^{52}$Co is more bound than predicted by the atomic mass evaluation. On the other hand, our experimental value for the $^{52}$Co ground state is significantly higher than the estimation based on mirror symmetry and known $\beta$-delayed proton data, $-34490(88)$ keV~\cite{Tu2016}.

\subsection{Excitation energies for the isomer $^{52}$Co$^m$ and the $T=2$ IAS in $^{52}$Co \label{sec:52coIAS}}
Based on the ground and isomeric state masses measured in the experiment, an excitation energy of $E_x=374(13)$ keV was determined for the isomer $^{52}$Co$^m$. This agrees well with the extrapolated value in the NUBASE 2012 evaluation, $E_x=380(100)$\# keV \cite{NUBASE12} which is simply taken from $E_x=377.749(5)$ keV \cite{NDS52-2015} for the analog state in the mirror nucleus $^{52}$Mn.

\begin{figure*}[!]
\centering
\includegraphics[width=0.8\textwidth,clip]{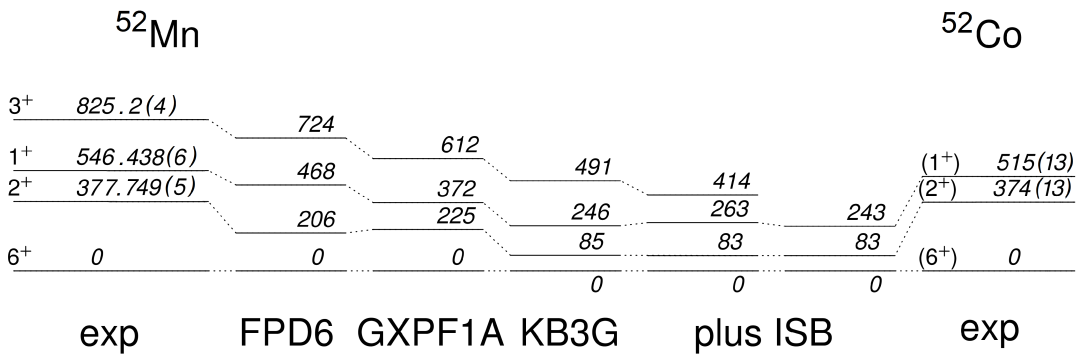}
\caption{Lowest levels observed in $^{52}$Mn and $^{52}$Co in comparison with large-scale shell-model calculations using the full $fp$ shell, $t=7$. Shown are the results obtained with FPD6, GXPF1A and KB3G without isospin-symmetry breaking terms (ISB), and with ISB for KB3G. \label{fig:shellmodel}}
\end{figure*}

The spins and parities for the lowest states in $^{52}$Co have not been experimentally verified. Thus, we performed large-scale shell-model calculations with the full $fp$ shell ($t=7$) to study the lowest levels in $^{52}$Co. The calculations were performed without isospin-symmetry breaking terms (ISB) for FPD6, GXPF1A and KB3G, as well as with ISB for KB3G (for details, see e.g. Ref.~\cite{Rudolph2008}). All calculations are in line with a $6^+$ ground state and a $2^+$ first excited (isomeric) state (see Fig.~\ref{fig:shellmodel}). However, all models have problems with producing the observed energy split between the $6^+$ and $2^+$ states. On the other hand, the experimental mirror energy differences between the $2^+$ and $1^+$ states in $^{52}$Co and $^{52}$Mn determined in this work,  $-4(13)$~keV and $-31(13)$~keV, respectively, are in good agreement with the KB3G calculations including ISB terms, which yield differences of zero keV and $-20$ keV, respectively.

\begin{figure*}
\centering
\includegraphics[width=0.68\textwidth]{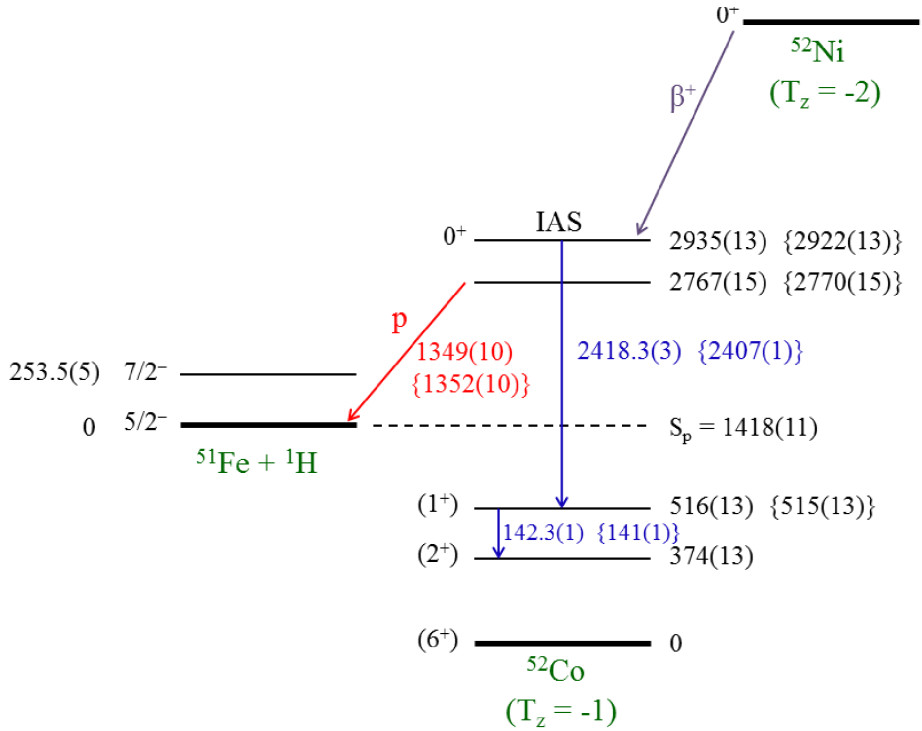}
\caption{\label{fig:52Co_levels} (Color online) Partial decay scheme for the $\beta^+$ decay of $^{52}$Ni with the proton separation energy $S_p$($^{52}$Co) and the excitation energy for the isomeric state $2^+$ from this work. All energies are in keV. The energies for the $\gamma$ transitions (shown in blue) and the proton energy $E_p$ are taken from Ref.~\cite{Dossat2007} and \cite{Orrigo2016} (in curly brackets). The energies for the $1^+$ state and the IAS are based on our value of $E_x$($^{52}$Co$^m$) and the $\gamma$-transitions from Refs.~\cite{Dossat2007} and \cite{Orrigo2016} (in curly brackets). The parameters of levels in $^{51}$Fe are taken from Ref.~\cite{NDS51-2006}. The proton line highlighted in red was previously thought to originate from the IAS but this work has shown it comes from a state lower than the IAS.}
\end{figure*}

The excitation energy of $^{52}$Co$^m$ has important consequences for studying the $T=2$ isobaric multiplet at $A=52$. Namely, the excitation energy of the $T=2$ IAS in $^{52}$Co can be determined based on the $\gamma$-$\gamma$ cascade from the $T=2$ IAS to the isomer \cite{Dossat2007}: $E_{IAS}(^{52}$Co$)=E_{\gamma 1}$(2418.3(3) keV)+$E_{\gamma 2}$(142.3(1) keV)+$E(^{52}$Co$^m)=2934(13)$ keV. However, a recent experiment performed at GANIL \cite{Orrigo2016} obtained a significantly different energy for the most intense gamma transition, $E_{\gamma 1}=2407(1)$ keV, and a slightly smaller energy for the second transition, $E_{\gamma 2}=141(1)$ keV. With these values, $E_{IAS}(^{52}$Co$)=2922(13)$ keV is obtained (see Fig.~\ref{fig:52Co_levels}). For both cases, the $\gamma$-$\gamma$ cascade is taken to feed the $(2^+)$ isomer, i.e. $^{52}$Co$^m$.

On the other hand, the energy for the $T=2$ IAS in $^{52}$Co can be determined based on the proton separation energy of $^{52}$Co and $\beta$-delayed protons observed from the IAS with center-of-mass energies of $E_{p,CM}=1349(10)$ keV \cite{Dossat2007} and $E_{p,CM}=1352(10)$ keV \cite{Orrigo2016}. Our new ground-state mass for $^{52}$Co results in a proton separation energy $S_ p(^{52}$Co$)=1418(11)$ keV, when the mass values for $^{51}$Fe and $^1$H are taken from Ref.~\cite{AME12}. Assuming the observed protons come from the IAS, the excitation energy should be $E_{IAS}(^{52}$Co$)=E_{p,CM}+S_p(^{52}$Co$)=2767(15)$~keV~\cite{Dossat2007} or 2770(15)~keV~\cite{Orrigo2016}. Thus, the obtained excitation energy is 167(20) keV \cite{Dossat2007} or 152(20) keV \cite{Orrigo2016} lower than that obtained via the $\gamma$-$\gamma$ cascade. 

The difference of around $150$~keV between the excitation energies of the IAS is much smaller than the excitation energy of the first excited state in $^{51}$Fe, $E_x=253.5(5)$ keV \cite{NDS51-2006}. A missed $\gamma$-transition of around 150 keV in $^{51}$Fe is also unlikely as it should have been observed in coincidence with the intense proton peaks. The discrepancy could be explained if the mass of $^{51}$Fe was around 150 keV off in AME12 \cite{AME12}. However, it is known with a precision of 9 keV and is based on two independent measurements \cite{Mueller1977,Tu2011}. Hence, the observed protons are most likely emitted from a state below the IAS (see~Fig.~\ref{fig:52Co_levels}). 

Thus, we obtain $E_{IAS}$($^{52}$Co)=2934(13) keV \cite{Dossat2007} for the excitation energy of the IAS in $^{52}$Co, or $E_{IAS}$($^{52}$Co)=2922(13) keV, if more recent values from Orrigo \emph{et al.} \cite{Orrigo2016} are used. Both values are much closer to the excitation energy of the $T=2$, $0^+$ mirror state in $^{52}$Mn, $E_x=2926.0(5)$ keV than what was obtained from the proton radioactivity data. The difference between Refs.~\cite{Dossat2007} and \cite{Orrigo2016} is mainly due to the discrepancy between the observed $\gamma$-transition energies, 2418.3(3) keV \cite{Dossat2007} and 2407(1) keV~\cite{Orrigo2016}. There is no clear explanation for the difference and a new measurement of the $\gamma$-$\gamma$ cascade would be required to obtain a more accurate excitation energy for the IAS.

\subsection{The masses of $^{52}$Fe and $^{52}$Fe$^m$\label{sec:52fe}}
\begin{figure*}[tbh]
\centering
\includegraphics[width=0.95\textwidth,clip]{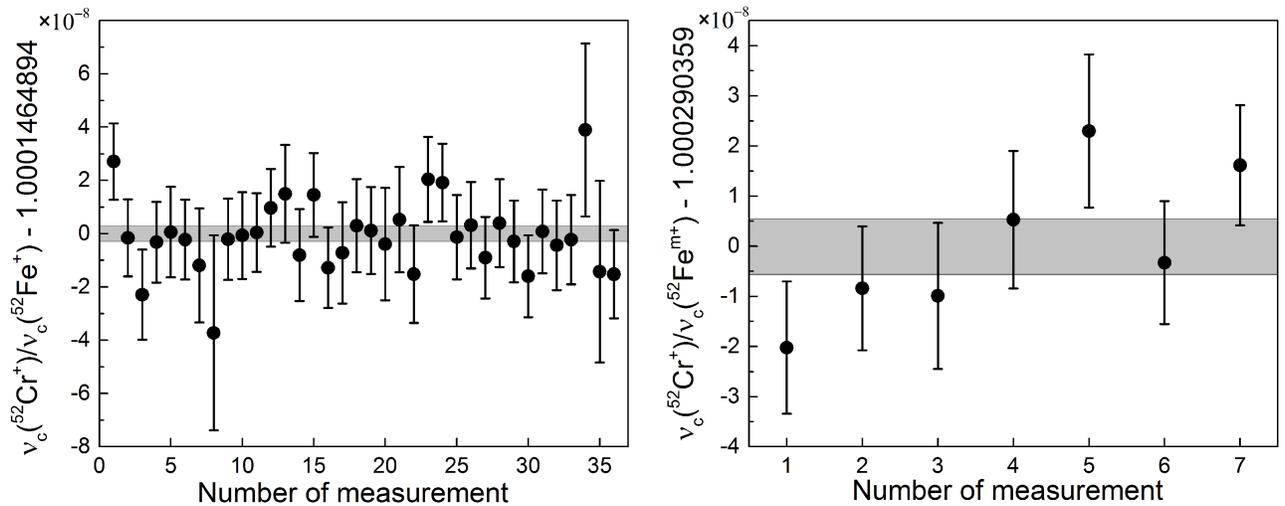}
\caption{Cyclotron-frequency ratios measured in this work for $^{52}$Fe$^+$ (left) and $^{52}$Fe$^{m+}$ (right). The gray-shaded bands represent the total uncertainty of the averaged frequency ratios.\label{fig:r52fe}}
\end{figure*}

\begin{figure}[tbh]
\centering
\includegraphics[width=0.45\textwidth,clip]{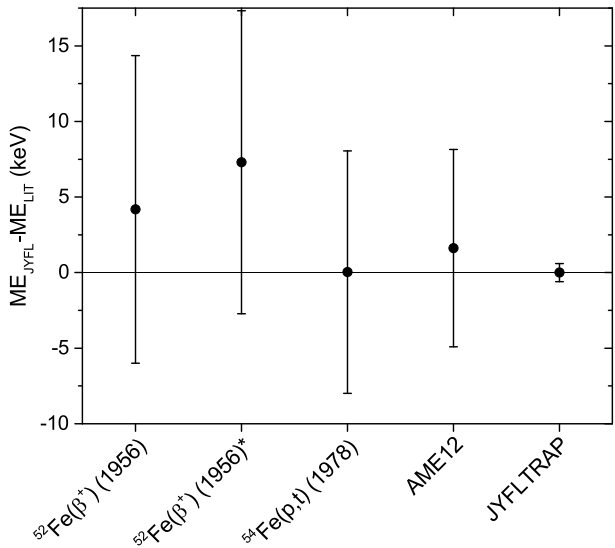}
\caption{Comparison of previous mass-excess values of $^{52}$Fe to the JYFLTRAP value determined in this work. The AME12 evaluation \cite{AME12part1} is mainly based on the $\beta^+$ decay of $^{52}$Fe \cite{Arbman1956} and the $^{54}$Fe$(p,t)^{52}$Fe reaction $Q$ value \cite{Kouzes1978}. For the $\beta$-decay value denoted by an asterisk, the JYFLTRAP value of $^{52}$Mn has been used. \label{fig:fe52comparison}}
\end{figure}

Altogether 36 cyclotron frequency ratios for the ground state $^{52}$Fe and seven ratios for the $12^+$ isomeric state $^{52}$Fe$^m$ were determined (see Fig.~\ref{fig:r52fe}). The weighted means of the frequency ratios, $1.0001464894(28)$ and $1.0002903590(56)$, yield atomic mass-excess values of $-48330.67(60)$ keV and $-41370.01(65)$ keV for the ground state and isomer, respectively. These are in good agreement with the literature values \cite{AME12}. Previously, the ground-state mass of $^{52}$Fe has been mainly based on the $\beta^+$ decay of $^{52}$Fe and the $^{54}$Fe$(p,t)^{52}$Fe reaction $Q$ value (see Fig.~\ref{fig:fe52comparison}). The excitation energy for the isomer, $E_x= 6960.7(9)$ keV, determined in this work differs by 2.7(10) keV from the literature value $E_x=6958.0(4)$ keV \cite{NDS52-2015} based on the observation of an $E4$ $\gamma$ transitions from this $12^+$ yrast trap in $^{52}$Fe \cite{Gadea2005}. It should be noted that the relative uncertainties of the measured frequency ratios, $2.8\times10^{-9}$ and $5.6\times10^{-9}$ for $^{52}$Fe$^+$ and $^{52}$Fe$^{m+}$, respectively, are much smaller than the relative uncertainty in the reference mass $\delta m/m(^{52}$Cr$)=1.2\times10^{-8}$. Therefore, the precision in the determined mass-excess values could be further improved via a more precise measurement of the reference $^{52}$Cr.

\subsection{Proton separation energy of $^{53}$Co and the energy of the protons emitted from $^{53}$Co$^m$\label{sec:Sp53Co}}
The mass of $^{52}$Fe is relevant for determining the proton separation energy of $^{53}$Co and, in particular, for the energy of the protons emitted from the $19/2^-$ high-spin isomer $^{53}$Co$^m$ \cite{Jackson1970,Cerny1970,Cerny1972}. By combining the newly measured $^{52}$Fe mass-excess value with the earlier JYFLTRAP mass measurements of $^{53}$Co ($-42657.3(15)$ keV \cite{Kankainen2010a}) and $^{53}$Co$^m$ ($-39482.9(16)$ keV \cite{Kankainen2010a}), a proton separation energy $S_p(^{53}$Co)=1615.6(16) keV and a center-of-mass energy $E_{p,CM}(^{53}$Co$^m$)=1558.8(17) keV for the protons from the high-spin isomer to the ground state of $^{52}$Fe is obtained. The proton separation energy is in a perfect agreement with the AME12 value $S_p(^{53}$Co)=1615(7) keV~\cite{AME12} but about four times more precise. Our value for the energy of the protons emitted from $^{53}$Co$^m$ is around 20 times more precise than obtained via proton-decay experiments $E_{p,LAB}=1530(40)$~keV \cite{Jackson1970}, $E_{p,LAB}=1570(30)$~keV \cite{Cerny1970}, and $E_{p,CM}=1590(30)$~keV \cite{Cerny1972}). Thus, we have demonstrated that Penning-trap measurements can provide precise calibration values for charged-particle spectroscopy.

\subsection{The mass of $^{52}$Mn\label{sec:52mn}}
\begin{figure}[tbh]
\centering
\includegraphics[width=0.45\textwidth,clip]{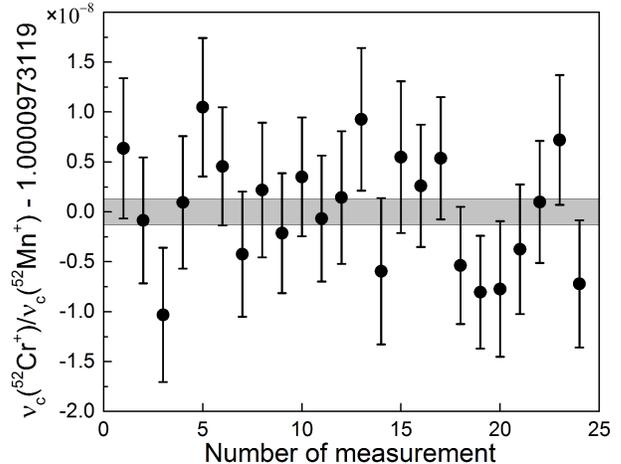}
\caption{Cyclotron-frequency ratios measured in this work for $^{52}$Mn$^+$. The gray-shaded band represents the total uncertainty of the averaged frequency ratio. \label{fig:r52mn}}
\end{figure}

\begin{figure}[tbh]
\centering
\includegraphics[width=0.45\textwidth,clip]{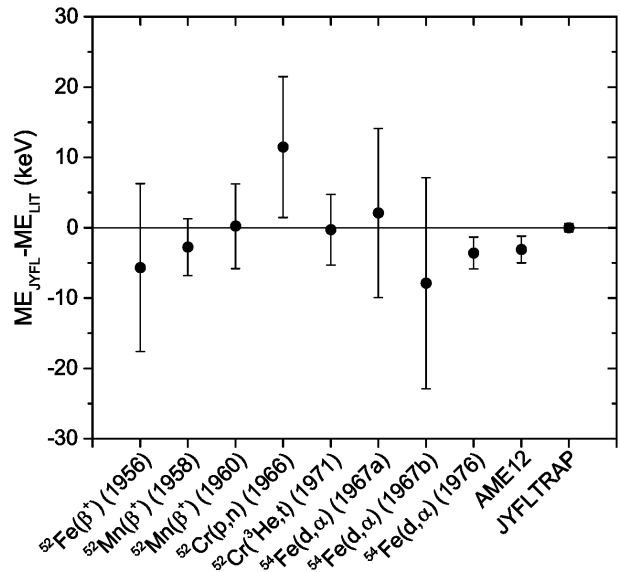}
\caption{Comparison of $^{52}$Mn mass-excess values from previous works and the AME12 \cite{AME12} to the JYFLTRAP value determined in this work. Previously, the mass of $^{52}$Mn has been studied via $\beta^+$ decays of $^{52}$Fe \cite{Arbman1956} and $^{52}$Mn $\beta^+$ decay \cite{Konijn1958,Katoh1960}, $^{52}$Cr$(p,n)$ \cite{Rickards1966}, $^{52}$Cr$(^3$He$,t)$ \cite{Becchetti1971}, as well as via $^{54}$Fe$(d,\alpha)$ reactions (see Refs.~\cite{Hansen1967} (1967a), \cite{Sperduto1967} (1967b), and \cite{Jolivette1976}.) \label{fig:mn52comparison}}
\end{figure}

A frequency ratio $r=1.0000973119(13)$ was derived as a weighted mean of 24 individual cyclotron frequency ratios for $^{52}$Mn$^+$ (see Fig.~\ref{fig:r52mn}). This yields an atomic mass-excess value of $-50709.97(59)$ keV for $^{52}$Mn. The JYFLTRAP value is 3(2) keV higher than the AME12 value ($-50706.9(19)$ keV \cite{AME12}) but three times more precise. The AME12 value is mainly based on a $Q$-value measurement of $^{54}$Fe$(d,\alpha)^{52}$Mn \cite{Jolivette1976} which gives a mass-excess value of $-50706.4(23)$ keV. In fact, $^{52}$Mn is an example of a nuclide close to stability whose mass has not been determined with modern techniques. The observed difference to the AME12 value shows that it is worthwhile to check such mass values which are based on measurements performed decades ago. Although our new value disagrees with Ref.~\cite{Jolivette1976}, it is in a rather good agreement with other, less precise experiments done on $^{52}$Mn (see Fig.~\ref{fig:mn52comparison}). Of these, the $\beta^+$ decay of $^{52}$Mn \cite{Katoh1960} and the value based on the $^{52}$Cr$(^3$He$,t)$ reaction \cite{Becchetti1971} agree very well with the present value. A more precise mass value for the reference $^{52}$Cr would be beneficial for $^{52}$Mn as well since the relative uncertainty of the frequency ratio, $1.3\times10^{-9}$, is nine times smaller than the relative uncertainty of the reference mass.

\subsection{The IMME for the $T=2$ quintet at $A=52$\label{sec:IMME}}
The IMME was studied for the $T=2$ quintet at $A=52$ using the mass values for $^{52}$Co$^m$, $^{52}$Fe, and $^{52}$Mn determined in this work together with the mass values of $^{52}$Cr and $^{52}$Ni adopted from AME12 \cite{AME12} as summarized in Table~\ref{tab:imme}. Our mass-excess value for the isomer $^{52}$Co$^m$ and the energies of the $\gamma$-$\gamma$ cascade observed from the IAS in Ref.~\cite{Orrigo2016} (see Sect.~\ref{sec:52coIAS}) were used for the mass excess of the IAS in $^{52}$Co. The excitation energy of the $T=2$ IAS in $^{52}$Fe, 8561(5) keV \cite{NDS52-2015}, is based on a study employing the $^{54}$Fe$(p,t)$ reaction \cite{Decowski1978}, where a doublet of two $0^+$ levels separated by around 4 keV were observed. For $^{52}$Mn, the IAS at 2926.0(5) keV \cite{NDS52-2015} has been identified in many experiments with the main contribution coming from a $^{52}$Cr$(p,n\gamma)$ study \cite{DelVecchio1973} where 2379.5(5) keV $\gamma$-rays from the IAS to the $1^+$ state at 546.438(6) keV were observed.

\begin{table}%[H] add [H] placement to break table across pages
\caption{Excitation energies, $E_{x,IAS}$, and mass-excess values, $ME_{IAS}$, for the $J^\pi=0^+,T=2$ isobaric analog states at $A=52$. For $^{52}$Fe and $^{52}$Mn, the mass-excess values for the IAS are based on the mass-excess values from this work and excitation energies from Ref.~\cite{NDS52-2015}. For $^{52}$Co, the mass excess for the IAS is based on the mass of $^{52}$Co$^m$ measured in this work and the $\gamma$-ray energies from Ref.~\cite{Orrigo2016}.\label{tab:imme}}
\begin{ruledtabular}
\begin{tabular}{llll}
Nuclide & $T_Z$ & $E_{x,IAS}$ (keV) & $ME_{IAS}$ (keV) \\
% Lines of table here ending with \\
\hline
$^{52}$Ni & -2 &	0 												&	-23474(700)\# \cite{AME12}\\
$^{52}$Co & -1 &	2922(13) 									&	-31410(11) \\
$^{52}$Fe &	0 &	8561(5) \cite{NDS52-2015} 	&	-39769.7(50)\\
$^{52}$Mn &	1 &	2926.0(5) \cite{NDS52-2015} &	-47783.97(77)\\
$^{52}$Cr &	2 &	0 													&	-55418.1(6) \cite{AME12}\\
\end{tabular}
\end{ruledtabular}
\end{table}

The results for the error-weighted quadratic and cubic fits for the IMME are given in Table~\ref{tab:immefits}. The reduced $\chi^2$ of 2.4 for the quadratic fit is well above one. However, the cubic coefficient $d=6.0(32)$ keV is within the $\pm 3\sigma$ limit from zero and, thus, compatible with zero. We checked the quadratic fit also without $^{52}$Ni, which is based only on the extrapolation of the mass surface \cite{AME12}. A slightly higher reduced $\chi^2$ value, $\chi^2/n=3.3$, and a cubic coefficient of $d=5.8(32)$ keV were obtained in this instance. If a quartic term $eT_Z^4$ is assumed instead of a cubic term, a coefficient $e=2.9(18)$ keV is obtained, again consistent with zero.

\begin{figure}[tbh]
\centering
\includegraphics[width=0.45\textwidth,clip]{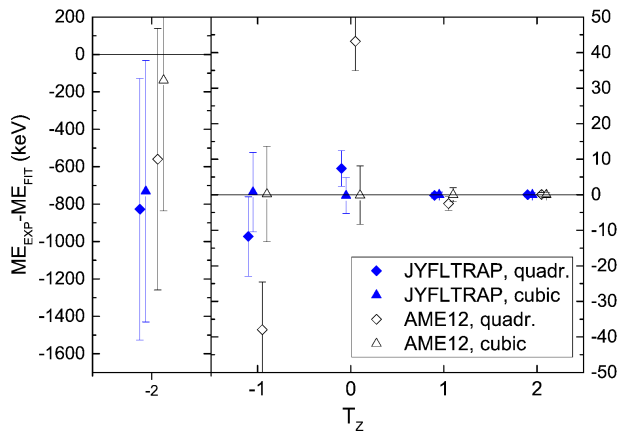}
\caption{(Color online) Differences of the mass-excess values of the $T=2$ quintet at $A=52$ to the quadratic and cubic fits from this work (see Table~\ref{tab:immefits}) or from AME12, using the masses of $^{51}$Fe and $^{1}$H together with the $E_{p,CM}=1349(10)$ keV from \cite{Dossat2007} for the IAS in $^{52}$Co.\label{fig:immediff}}
\end{figure}

Previously, a non-zero coefficient $d=28.8(45)$ keV and unstable behavior of coefficient $c$ from quadratic to cubic fits were observed \cite{MacCormick2014}. We can now confirm that these have been due to erroneous data used in the fits. Prior to this experiment, it was assumed that protons with $E_{p,CM}=1352(10)$ keV \cite{Orrigo2016} originate from the $T=2$ IAS in $^{52}$Co. Using the mass-excess values of $^{51}$Fe and $^{1}$H from AME12 \cite{AME12} together with the proton energy, a mass-excess value of $-31561(13)$ keV for the IAS in $^{52}$Co is obtained. This differs by 152(17) keV from the value determined in this work. For comparison, we performed similar IMME fits using the AME12 data for the ground-state masses and $-31564(13)$ keV for the IAS in $^{52}$Co. The reduced $\chi^2$ for the quadratic fit was $18.9$ and the cubic coefficient $d=29.3(48)$ keV. Figure~\ref{fig:immediff} shows differences between the mass-excess values and the quadratic and cubic fits for the dataset determined in this work and with the AME12 values (assuming that the 1349-keV protons originate from the IAS). Clearly, a better agreement is achieved with our data. The fits of the new dataset also suggest that the mass-excess value for $^{52}$Ni might be higher, meaning it could be less bound than predicted in the AME12.

\begin{figure}[tbh]
\centering
\includegraphics[width=0.45\textwidth,clip]{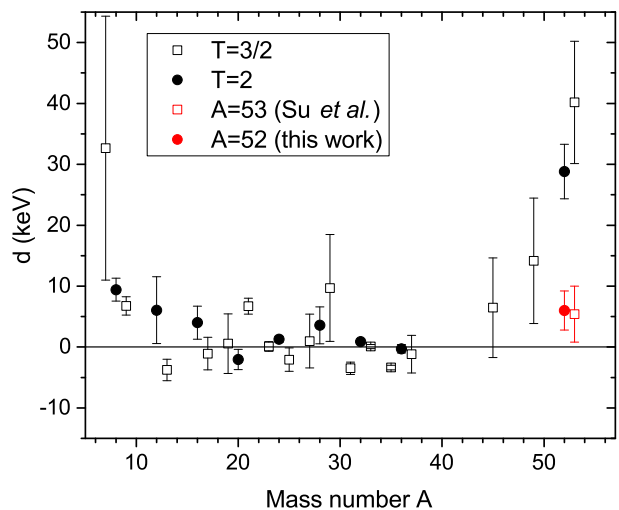}
\caption{(Color online) Cubic coefficients $d$ for the $T=3/2$ quartets and $T=2$ quintets from Ref.~\cite{MacCormick2014}, with $A=21$ and $A=31$ updated from recent publications \cite{Gallant2014} and \cite{Kankainen2016}. The cubic coefficient observed in this work for the $A=52$ quintet, $d=6.0(32)$ keV, is close to the value obtained recently for the $A=53$ quartet, $d=5.4(46)$ keV by Su \emph{et al.}~\cite{Su2016}, and consistent with zero. These new values are highlighted in red. \label{fig:immecoeff}}
\end{figure}

The cubic coefficients for the $T=3/2$ quartets and $T=2$ quintets have been plotted in Fig.~\ref{fig:immecoeff}. Earlier, a trend of increasing cubic coefficients after entering the $f_{7/2}$ shell has been observed. However, a recent observation of $\gamma$-rays from the IAS in $^{53}$Co following the $\beta$ decay of $^{53}$Ni showed that the IAS is lower than anticipated by $\beta$-delayed proton data from Ref.~\cite{Dossat2007}. With the new excitation energy, the cubic coefficient for the $A=53$ quartet is $d=5.4(46)$ keV \cite{Su2016}, and thus does not suggest a breakdown in the IMME. In this work, we obtained a very similar cubic coefficient for the $T=2$ quintet at $A=52$, $d=6.0(32)$ keV, and confirmed that the intensive beta-delayed proton group observed in the beta-decay of $^{52}$Ni \cite{Orrigo2016,Dossat2007} does not originate from the IAS in $^{52}$Co but from a state below it. This is also understandable since the beta-delayed protons from the $T=2$ isobaric analogue state decaying to the ground state of $^{51}$Fe ($T=1/2$) are isospin-forbidden, and thus, the proton-branch from the IAS should be rather small as it is possible only via isospin mixing between the $T=1$ and $T=2$ states in $^{52}$Co or $T=1/2$ and $T=3/2$ states in $^{51}$Fe, respectively. 

In conclusion, there is no evident change in the cubic coefficients after entering the $f_{7/2}$ shell. Although the coefficients are on the order of some keV, they are still compatible with zero within the $\pm 3\sigma$ limit. The mass determination of the most exotic member of the $T=2$ multiplet at $A=52$, $^{52}$Ni, would be crucial to provide a more stringent test of the IMME at the heavier mass region. 

\begin{table}%[H] add [H] placement to break table across pages
\caption{Coefficients and the reduced $\chi^2$ values for the quadratic and cubic IMME fits (in keV) for the $T=2$ quintet at $A=52$.\label{tab:immefits}}
\begin{ruledtabular}
\begin{tabular}{lll}
&	Quadratic	& Cubic\\
% Lines of table here ending with \\
\hline
a &	-39777.1(30) 	&	-39769.4(50)\\
b	& -8192.9(46)		& -8192.8(46)\\
c	& 186.2(16)			& 172.2(75)\\
d	&								& 6.0(32)\\
$\chi^2/n$&	2.4	& 1.1\\
\end{tabular}
\end{ruledtabular}
\end{table}

\subsection{Implications for the rapid proton capture process\label{sec:rp}}
The $rp$ process proceeds along nuclei close to the $N=Z$ line mainly via proton captures and $\beta^+$ decays, resulting in a thermonuclear runaway and a sudden release in energy observed, for example, in type I X-ray bursts \cite{Wallace1981,Schatz1998}. Proton-capture $Q$ values are essential for modeling the $rp$ process as they determine the path: the ratio of inverse photodisintegration reactions to the total proton-capture rate ($\lambda_{\gamma,p}/N_A\langle\sigma v\rangle$) depends exponentially on the proton-capture $Q$ values. Those can have a significant effect as demonstrated in Ref.~\cite{Parikh2009}. 

With the new mass-excess value determined in this work for $^{52}$Co, proton-capture $Q$ values for $^{51}$Fe$(p,\gamma)^{52}$Co and $^{52}$Co$(p,\gamma)^{53}$Ni can be experimentally determined for the first time. The new values, $Q(^{51}$Fe$(p,\gamma)^{52}$Co)=1418(11) keV and $Q(^{52}$Co$(p,\gamma)^{53}$Ni)=2588(26) keV, differ significantly from the extrapolated values of 1077(196)\# keV and 2930(197)\# keV \cite{AME12}, respectively. In other words, $^{52}$Co is around 340 keV more proton bound and $^{53}$Ni less proton bound than expected from the extrapolations of the mass surface in AME12. 

\begin{figure}[tbh]
\centering
\begin{tabular}{c}
\includegraphics[width=0.45\textwidth,clip]{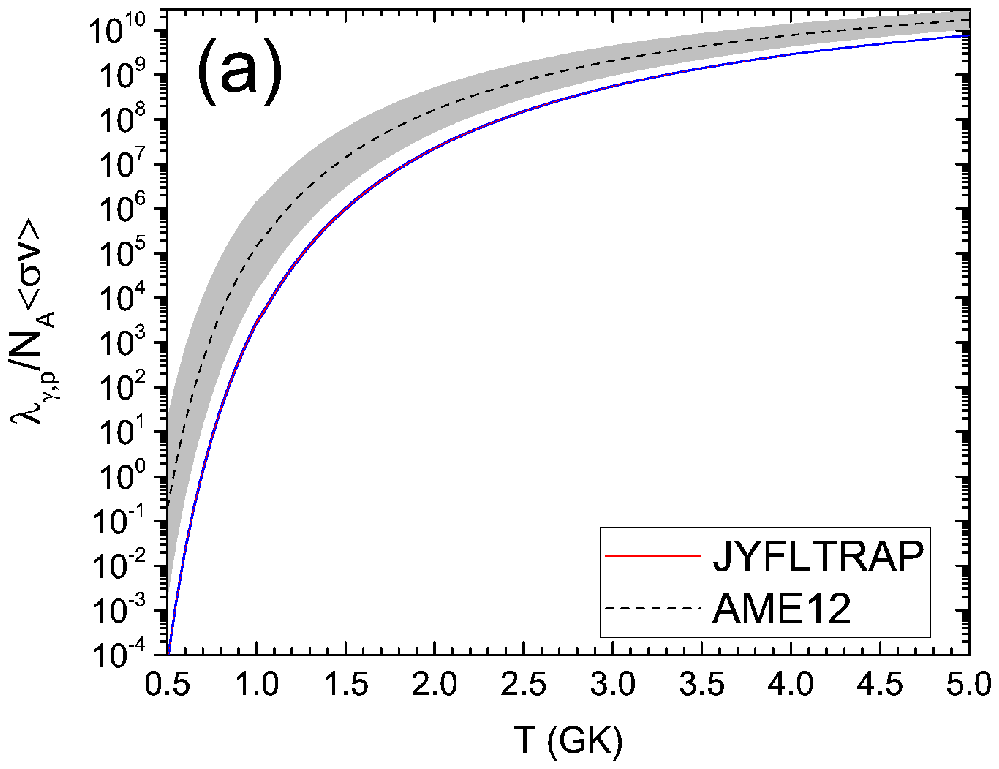}\\
\includegraphics[width=0.45\textwidth,clip]{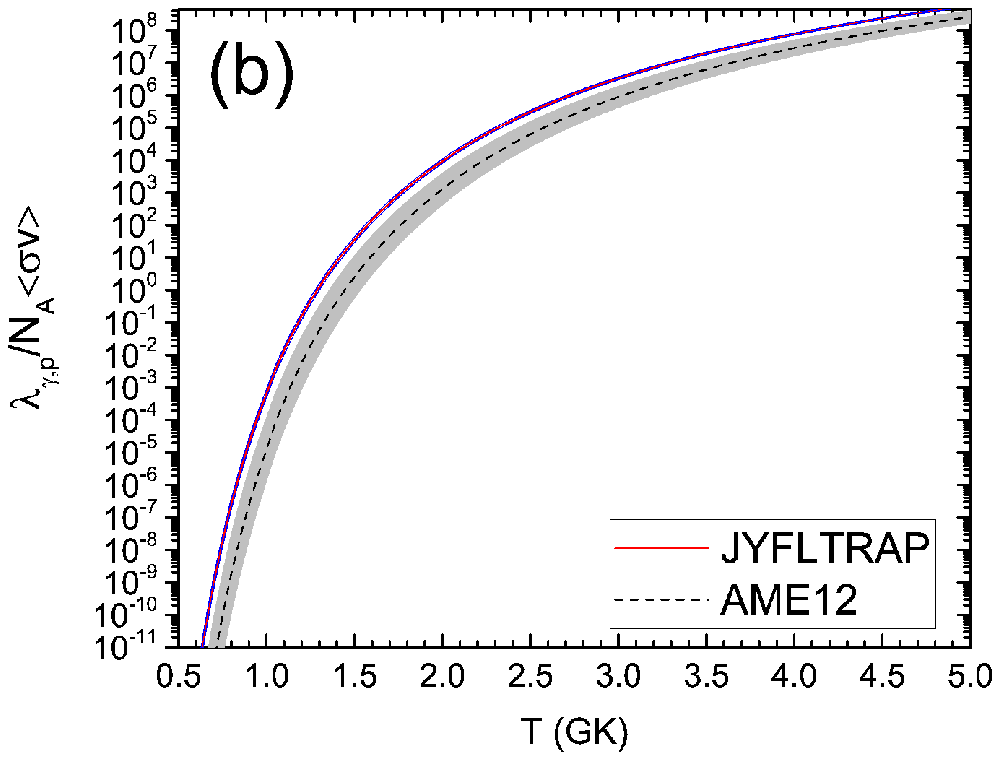}\\
\end{tabular}
\caption{(Color online) Ratio of the photodisintegration $(\gamma,p)$ to the proton-capture rate $N_A\langle\sigma v\rangle$ for (a) $^{51}$Fe$(p,\gamma)^{52}$Co - $^{52}$Co$(\gamma,p)^{51}$Fe and (b) $^{52}$Co$(p,\gamma)^{53}$Ni-$^{53}$Ni$(\gamma,p)^{52}$Co reactions. The gray-shaded regions show the uncertainty band related to the AME12 $Q$ value. The $Q$-value related uncertainties for the JYFLTRAP results are invisible on this scale.\label{fig:rp}}
\end{figure}

In Fig.~\ref{fig:rp} we show the effect of the new $Q$ values on the photodisintegration versus proton capture rate ratio. With the new $Q$ values, the route via $^{52}$Co is more likely than before as $^{52}$Co is more proton-bound. For example, when the experimental $Q$ value is used instead of the extrapolated AME12 value, photodisintegration rates on $^{52}$Co are suppressed by a factor of around 50-3000 compared to the proton-capture rates on $^{51}$Fe at temperatures below 1 GK. Although more detailed $rp$-process calculations would be needed to find out the effect on the whole $rp$ process, the big change in the $^{52}$Co mass value significantly changes the calculations related to the proton captures and photodisintegration reactions involving $^{52}$Co. 

\section{Conclusions and Outlook\label{sec:conclusions}}
In this work, we have performed direct mass measurements of $^{52}$Co, $^{52}$Co$^m$, $^{52}$Fe, $^{52}$Fe$^m$, and $^{52}$Mn with the JYFLTRAP double Penning-trap mass spectrometer. The masses of $^{52}$Co and its isomer $^{52}$Co$^m$ have been experimentally determined for the first time. The new mass value for $^{52}$Co is significantly lower than that obtained via extrapolations in the AME12 showing that it is more bound than expected. The obtained excitation energy of the isomer, $E_x=374(13)$ keV, is in good agreement with its analog state in $^{52}$Mn with $E_x=377.749(5)$ keV \cite{NDS52-2015}. Based on isospin symmetry and supported by shell-model calculations, we assume $I^\pi=2^+$ for the isomeric state.

An important consequence of the mass measurements of the $^{52}$Co ground and isomeric states is that the mass for the $T=2$ IAS can be determined using data from $\beta$ decay of $^{52}$Ni \cite{Dossat2007,Orrigo2016}. We have found that the protons assumed to originate from the IAS in Ref.~\cite{Dossat2007} must come from a state at around 2770(15) keV in $^{52}$Co, which is significantly lower than the excitation energy determined for the IAS in this work, $E_x=2922(13)$ keV, based on the observed $\gamma$-$\gamma$ cascade \cite{Dossat2007,Orrigo2016} from the IAS to the $(2^+)$ isomeric state in $^{52}$Co. The new excitation energy for the IAS agrees well with the analogue state in the mirror nucleus $^{52}$Mn, 2926.0(5) keV. It is interesting that the IAS seems to decay only via $\gamma$ transitions since the proton decays are isospin-forbidden, whereas the state below it has a substantial proton branch but no observed $\gamma$ transitions. In the future, further experiments to confirm the state from which the observed protons come from are needed. In addition, the discrepancy between the measured $\gamma$-transition energies of 2418.3(3) keV \cite{Dossat2007} and 2407(1) keV \cite{Orrigo2016} should be studied to improve the accuracy of the $T=2$ IAS in $^{52}$Co. 

The masses of $^{52}$Fe, $^{52}$Fe$^m$, and $^{52}$Mn have been determined with a much higher accuracy than in AME12. The precision in the $^{52}$Fe mass value has been improved by a factor of around twelve, which allows a precise determination of the proton separation energy of $^{53}$Co, $S_p(^{53}$Co)=1615.6(16) keV. In addition, the energy of the protons emitted from the high-spin isomer in $^{53}$Co$^m$ to the ground state of $^{52}$Fe has been determined with unprecedented precision, $E_{p,CM}=1558.8(17)$ keV. Penning-trap measurements of $^{52}$Fe, $^{53}$Co, and $^{53}$Co$^m$ at JYFLTRAP have therefore delivered an external calibration value for proton-decay experiments. 

Whereas the masses of $^{52}$Fe and $^{52}$Fe$^m$ agree well with AME12, $^{52}$Mn shows a deviation of $-3(2)$ keV. The observed deviation demonstrates the importance of measuring masses also closer to stability. Previous experiments may have been performed already several decades ago, and the mass accuracy can be improved considerably using Penning-trap mass spectrometry. It should be noted that a more accurate mass value for the reference $^{52}$Cr would improve the precision of the mass values for $^{52}$Fe, $^{52}$Fe$^m$, and $^{52}$Mn determined in this work. Presently, the mass of $^{52}$Cr is linked to $^{55}$Mn via $^{52}$Cr$(n,\gamma)^{53}$Cr$(n,\gamma)^{54}$Cr$(p,\gamma)^{55}$Mn \cite{AME12part1}, where $^{55}$Mn has been measured with respect to $^{85}$Rb at ISOLTRAP \cite{Naimi2012}.

The first mass measurement of $^{52}$Co provides also first experimental proton separation energies for $^{52}$Co and $^{53}$Ni, 1418(11) keV and 2588(26) keV, respectively. These are also the proton-capture $Q$ values for the proton captures $^{51}$Fe$(p,\gamma)^{52}$Co and $^{52}$Co$(p,\gamma)^{53}$Ni, which affect $rp$-process calculations. Since $^{52}$Co has been found to be more bound than predicted in AME12, photodisintegration reactions on $^{52}$Co are not so dominant as previously predicted, thus making it more likely that the $rp$ process proceeds via $^{51}$Fe$(p,\gamma)^{52}$Co.

Finally, we have thoroughly studied the IMME for the $T=2$ quintet at $A = 52$ using the new mass values determined in this work. The quadratic fit results in $\chi^2/n=2.4$, which corresponds to around 10~\% probability that the quintet can be described with a parabola. However, the cubic coefficient, $d=6.0(32)$ keV does not support a breakdown in the IMME. The cubic coefficient is significantly lower than obtained in the previous IMME evaluation, $d = 28.8(45)$ keV \cite{MacCormick2014}, and close to the value recently determined for the $T=3/2$ quartet at $A=53$ \cite{Su2016}. The new value does not suggest a trend of increasing cubic coefficients when entering the $f_{7/2}$ shell. In the future, a mass measurement of $^{52}$Ni would provide a possibility for a more stringent test of the IMME at $A=52$.

\begin{acknowledgments}
This work has been supported by the Academy of Finland under the Finnish Centre of Excellence Programme 2012–2017 (Nuclear and Accelerator Based Physics Research at JYFL) and the Swedish Research Council (VR 2013-4271). A.K., D.N., and L.C. acknowledge support from the Academy of Finland under grant No. 275389.
\end{acknowledgments}

\bibliography{A52}

\end{document}